\documentclass[twocolumn]{paper}                  

\usepackage{graphicx}
\usepackage{amssymb,amsmath,multirow} 
\usepackage{rotating} \usepackage{setspace} \usepackage[T1]{fontenc}
\usepackage{url}
\usepackage{geometry}
\usepackage{mathpazo}
\usepackage{natbib}

\begin{document}
\onecolumn
\vspace{3cm}
\title{Cascades in real interbank markets }

{\small
\author{Fariba Karimi \\
Ume\r{a} University\\
 Department of Physics, Icelab \\
90187 Ume\r{a}, Sweden\\
fariba.karimi@physics.umu.se        
       \and
        Matthias Raddant \\
			Kiel Institute (IfW) and\\
							CAU Kiel, Department of Economics\\
							24118 Kiel, Germany\\
							raddant@economics.uni-kiel.de} 	
}
\maketitle

\vspace{1cm}
\begin{center}
{\itshape 
 The final publication is available at link.springer.com\\
http://link.springer.com/article/10.1007/s10614-014-9478-z\\
to be published in Computational Economics
}
\vspace{1cm}
\end{center}
\begin{abstract}
We analyze cascades of defaults in an interbank loan market. The novel feature of this study is that the network structure and the size distribution of banks are derived from empirical data. We find that the ability of a defaulted institution to start a cascade depends on an interplay of shock size and connectivity. Further results indicate that the ability to limit default risk by spreading the lending to many counterparts decreased with the financial crisis.
To evaluate the influence of the network structure on market stability, we compare the simulated cascades from the empirical network with results from different randomized network models. The results show that the empirical network has non-random features, which cannot be captured by rewired networks. The analysis also reveals that simulations assuming homogeneity for size of banks and loan contracts dramatically overestimates the fragility of the interbank market.

\vspace{1cm}
\textbf{keywords:} interbank loan network -- systemic risk -- cascades -- null models

\end{abstract}
\thispagestyle{empty}

\twocolumn
\section{Introduction}

In this paper we investigate the likelihood of cascades of defaults in an interbank loan market. The novel feature of this investigation is that we use empirical data to calibrate our model and compare the results with different null models. This also allows us to make a statement about the accuracy of the existing literature on cascades in financial markets. We find that stylized random network models are very likely to overestimate cascade effects after a shock to the system. 

Overnight loans are a financial instrument to redistribute liquidity between banks in the short term. These loans are settled in two different ways. Most of them are settled as over-the-counter (OTC) agreements, about which data is not systematically collected. Studies on interbank loan markets therefor often rely on data that is reconstructed from the observation of payments between banks \citep[see, e.g.,][]{hejm}. The data that we analyze here is not from OTC transactions.  It comes from the Italian trading platform E-mid. In principle every bank could join this platform, and in fact roughly one third of the users are foreign banks. Banks can trade loans with different maturities and different currencies. In our study we focus on the euro-denominated (unsecured) overnight loans, which constitute by far the largest share of all transactions. Earlier studies have shown that the Italian market can be regarded as representative for the Euro money market, as the relative order flow to this market was relatively stable \citep[see, e.g.,][]{emidrep}.

There is an increase of research on linkages between banks since the 2008 financial crisis. Some European markets have been analyzed for example by \cite{furfine}, \cite{hartmann}, and \cite{cocco}. The U.S. money market system, Fed-wire, has been analyzed for example by \cite{asduffie} and \cite{kimmo}. The interest in these markets has risen for two reasons: the partial collapse of the interbank markets itself and the increasing need for risk assessment in the bank network in general. The contagious effects that played a big role in the events after the Lehman default showed that a micro-prudential analysis of banks' exposures to counterparts does not capture the systemic risks that the default of a bank can pose \citep[see, e.g.,][]{may,lux}.

Some earlier studies on the E-mid market exist \cite[see, e.g.,][]{ioriIT}. From the analysis of conditional trading volume \citep[][]{frickelinks} we know that no pronounced clusters of banks exist in the Italian market. The development of spreads and lending relationships have been analyzed by \cite{me}, who found that banks try hard to avoid large exposures since 2008. The network structure of the market \citep[see also][]{frickeemid} can at best be described as a core-periphery structure, similar to the findings by \cite{vonpeter} for the German market. In parallel to our study, default cascades in the E-mid market have also been analyzed by \cite{roukny}, although their focus is on the search for an optimal network structure. We can however confirm one of their findings, which is that network structure matters mainly in fragile markets. 

Several studies have tried to analyze systemic risk in interbank loan networks, for example by simulating knock-on effects in the case of a default. From relatively simple models like \cite{nier},  who study cascade effects in a Bernoulli random graph, research has evolved to more fine-tuned models like in \cite{cascadepnas}, where more realistic network structures and liquidity effects are taken into account. Mechanisms of loss amplification have been discussed by \cite{battgatti} and \cite{shin}. \cite{Boss} show for Austria that the network structure between banks is somewhat close to a scale-free graph, and that loan and bank size distributions can be approximated by a power law.  \cite{muller} analyzes cascades in the system of interbank loans for Switzerland, \cite{santos} analyze the Brazilian market, using a more detailed data set. \cite{mistrulli} shows how missing data on links can be approximated efficiently. The paper by \cite{eisenoe} is more on the theoretical side, the authors discuss cyclical interdependence and stability in a system of mutual repayments. A recent paper by \cite{glass} builds on the latter and analyzes how likely differently distributed shocks can lead to contagion. Our paper is methodologically close to the one by \cite{nier}, but uses empirical data on the loan network. The findings of \cite{glass} and also the findings in the aforementioned empirical studies hint at limitations of simulation studies that use homogeneously sized banks or degree. Specifically, it is not possible to characterize systemically important nodes or to create realistic chains of defaults if homogeneity is assumed. 

Cascade models certainly do not reflect all aspects of real financial markets. For the case of financial institutions, one problem is that we do not have (and probably never will have) data on the activity of banks for all financial products through which networks between them exist. The second problem is that we cannot know how the interbank market would continue on the days and weeks after a default, since in this kind of cascade simulations the start of the cascade is defined by the (hypothetical) removal of one node (bank) from the network. In this respect, the analysis of default cascades differs from studies on disease spreading or cascades in social networks \citep[see, e.g.,][]{holme,karimi} in which cascades with a temporal structure are analyzed. 

In this study we do not aim at a complete model of financial contagion. For this, more complex models with different layers of connections and a dynamic structure would be needed \citep[see][]{kok}. However, in order to build these multiplex models, certain layers have to be approximated and rules for the behavior of participants have to be calibrated. The results that we obtain from the analysis of one known layer, the overnight market, are an important input for this modeling, as our results can serve as a benchmark for calibration and help to asses the margin of error which different specifications of network models can have. 

For this reason, we set up an interbank network that is derived from empirical data on the Italian interbank loan market (E-mid). Specifically, we use daily snapshots of loan contracts in this market and derive a network of netted daily exposures. We randomly choose a bank to default and analyze knock-on effects to other market participants, based on the network given by interbank loans. We generate simplified balance sheets for each bank to determine its solvency. The balance sheet items that are not given by the data on interbank lending are hypothetical, but they are generated to be consistent with their interbank behavior, i.e., we obtain a network with a reasonable distribution of bank size, where all banks have the same a-priori ability (relative to their size) to absorb a shock. Finally we compare the results with those from different network models (null models), in which distributional properties of the network are simplified. This allows us to judge which kind of networks models should be used in cases where empirical data is not available.

The remainder of the paper goes like this: In Section \ref{sec:methods} we explain the data set, how we derive the bank's balance sheets, and how we simulate cascade effects. In Section \ref{sec:results} we discuss the simulation results. In Section \ref{sec:null} we compare the simulation results with those of different null models from random network. Section \ref{sec:conclusions} concludes the paper.

\section{Material and methods}\label{sec:methods}

\subsection{Loan data and bank balance sheets}

\begin{table*}
\begin{center}
\begin{tabular}{c|c c c c}

 & <nodes> & <links> & <degree> & <lending> \\
 \hline
\hline
2006  & 128 $\pm$9 & 355 $\pm$ 48 & 5.5 $\pm$ 0.5 & 20,953 $\pm$ 4,240 \\
2011  & 70 $\pm$8  & 161 $\pm$ 29 & 4.5 $\pm$ 0.5 & 4,261 $\pm$ 1,001 \\
\end{tabular}
\end{center}
\caption{Network properties}{\footnotesize The table shows the properties of the 257 daily networks for 2006 and 2011. We report the average number of links, the average degree, and the average lending volume together with their standard deviation. The lending volume is measured in millions in euro.}\label{tab:stats}
\end{table*}

For our investigation we use tick data from the interbank loan trading platform E-mid. In this data set the names of the banks have been replaced with an identifier, which allows us to infer the nationality of the bank, but not its name. For each transaction we observe the time, the traded amount, the interest rate, the identifier of the counterparts, and who was the initiator of a transaction. For our analysis we aggregate and net the loan contracts between all pairs of banks (nodes) and derive a directed weighted network for each day (see also Table \ref{tab:stats}). The weights of the links are given by the net amount of borrowing between the banks. The trading volume in the interbank market has changed considerably as a result of the financial crisis of 2008. Hence, for our investigation we choose two years in which the general characteristics of the network are comparably stable, the first one before and the second one after the outbreak of the crisis, 2006 and 2011.

In our assessment of the fragility of this network, we repeatedly choose one bank as the initial defaulter. We assume that this bank will not repay any of its interbank loans. Whether banks that lent to a defaulted counterpart default themselves, depends on the amount of capital (equity) of a bank, thus, the ability to write off lost interbank lending. We assume that no other balance sheet operations can be made to prevent a default.

\begin{table}[bt]
\begin{center}
\begin{tabular}{c|c}

 Assets & Liabilities \\
 \hline
\hline
Assets $A$ & Capital $C$ \\
        & Deposits $D$ \\
 Lending $L$ & Borrowing $B$ \\

\end{tabular}
\end{center}
\caption{Stylized bank balance sheet}{\footnotesize The interbank market activity of the bank shows as lending $L$ and borrowing $B$ in the balance sheet. In our simulation the remaining assets $A$ are exogenous. The amount of capital is always in a fixed proportion to total assets. Customer deposits $D$ conclude the liabilities side of the balance sheet. }\label{tab:bsheet}
\end{table}

Hence, while the amount of interbank loans are known exactly, we have to approximate the remaining items on the banks' balance sheets. The balance sheet contains external assets $A$, interbank lending $L$, capital $C$, deposits $D$, and interbank borrowing $B$ (see Table \ref{tab:bsheet}). We approximate the total assets for each bank by assuming that it is proportional to the average of the trading volume $TV = L + B$ of the last 10 days in which the bank was active in the market.\footnote{We choose this specification to ensure that bank's capital is always consistent with interbank market exposures and does not fluctuate wildly. We obtained qualitatively similar simulation results by estimating $TA$ only from the current day's activity, and for estimating $TA$ only from interbank lending. However, the results were more noisy, and, in the latter case, lead to an indeterminacy for banks that are only borrowing.} 
\begin{eqnarray}
<TV>_t = \frac{\sum_{i=0}^{9} (L_{t-i}+B_{t-i})}{10}
\end{eqnarray}
We assume that the value of all balance sheet items are constant fractions of the total assets $TA$. The parameter values $\theta$ and $\gamma$ that we choose for the baseline scenario are the same like in \cite{nier}. They are in line with the empirical findings of \cite{upper} and the figures reported in the stress-test of the \cite{eba}.  In the baseline scenario, the ratio of interbank lending $L$ to total assets $TA$ is set to $\theta = 0.2$ for each bank. Hence, the total assets $TA$ can be written as
\begin{eqnarray}
 TA = \frac{<TV>}{2 \theta}.
\end{eqnarray}
This already determines the asset side of the balance sheet, since the external assets are then given by $A=TA-L$.
The balance sheet identity implies that for each bank 
\begin{eqnarray} 
TA = A  + L = C + D + B .
\end{eqnarray}
The ratio of capital to total assets is set to $\gamma=0.05$, hence $C=\gamma TA$. Since $B$ is given by the data, this determines the liabilities side of the balance sheet:
\begin{eqnarray}
D = (1- \gamma) TA - B.
\end{eqnarray}

\subsection{The model}

For each day of our sample, we simulate the default of each bank that is active as a borrower. We think it is important to simulate the default of all banks, since focusing on only larger banks would bias the results towards the local properties of large banks. We assume that the initial default is triggered by an external cause, which is not in the scope of our paper.  As the consequence of this default, the bank will not meet its payment obligations on the interbank loan market. Hence, all banks which are net lenders to this one will suffer losses to the position $L$. Assuming that the share of a bank's exposure to the defaulting institution is $\phi$, the position $L$ will be reduced to $(1-\phi)L$. This loss has to be compensated by other positions of the banks balance sheet. Since we consider the short-run effects directly after a default, losses can only be absorbed by the capital position $C$. Summing up, this means that a bank survives a  shock as long as

\begin{eqnarray}
(1-\phi)L+ A  - B - D > 0.
\end{eqnarray} 

To check the solvency condition, we calculate the remaining amount of capital (equity) for each bank after the shock to the system. 
The simulation of defaults in this system takes place in several rounds. After the initial default of one bank, only other banks with an exposure to this bank (net lenders) will be effected. We have to recalculate the balance sheets of all these banks, and determine, if they could absorb the initial shock. Eventually some of them default themselves. In this case, we have to calculate the 2nd round effects on all banks that are net lenders to this institution. From this point on, also banks that were not connected to the initially defaulted bank can be effected. The recalculation of balance sheets is iteratively repeated until no further bank is found to default.

On the level of the individual bank we have to consider the shock that it might receive from a loan that is not repaid. Assume that bank $i$ is default, because it has received a shock $S_{\rightarrow i}$. It passes a residual shock to all its lenders $k_i$. The residual shock is given by the difference between $i$`s shock and its capital, $S_{\rightarrow i}-C_i$. Let us assume that $L_{ji}$ represents the amount that bank $j$ lent to bank $i$. $B_i$ represents the amount that $i$ borrowed from the set of banks $k_i$. In case that banks $i$'s residual shock is larger than its total borrowing, bank $j$ will receive a shock equal to the loaned amount $L_{ji}$. Otherwise bank $i$ will repay a fraction of the loan:

\begin{eqnarray}
\begin{aligned}
S_{ij} = L_{ji} \enskip \text{if}  \enskip S_{\rightarrow i} - C_i > B_i \\
S_{ij} = \frac{(S_{\rightarrow i}-C_i) L_{ji}}{\sum_k L_{ki}}  \enskip \text{if}  \enskip S_{\rightarrow i} - C_i < B_i
\end{aligned}
\end{eqnarray}

\begin{figure*}
\begin{center}
\begin{minipage}[b]{0.47\linewidth}
\centering
\includegraphics[width=\linewidth]{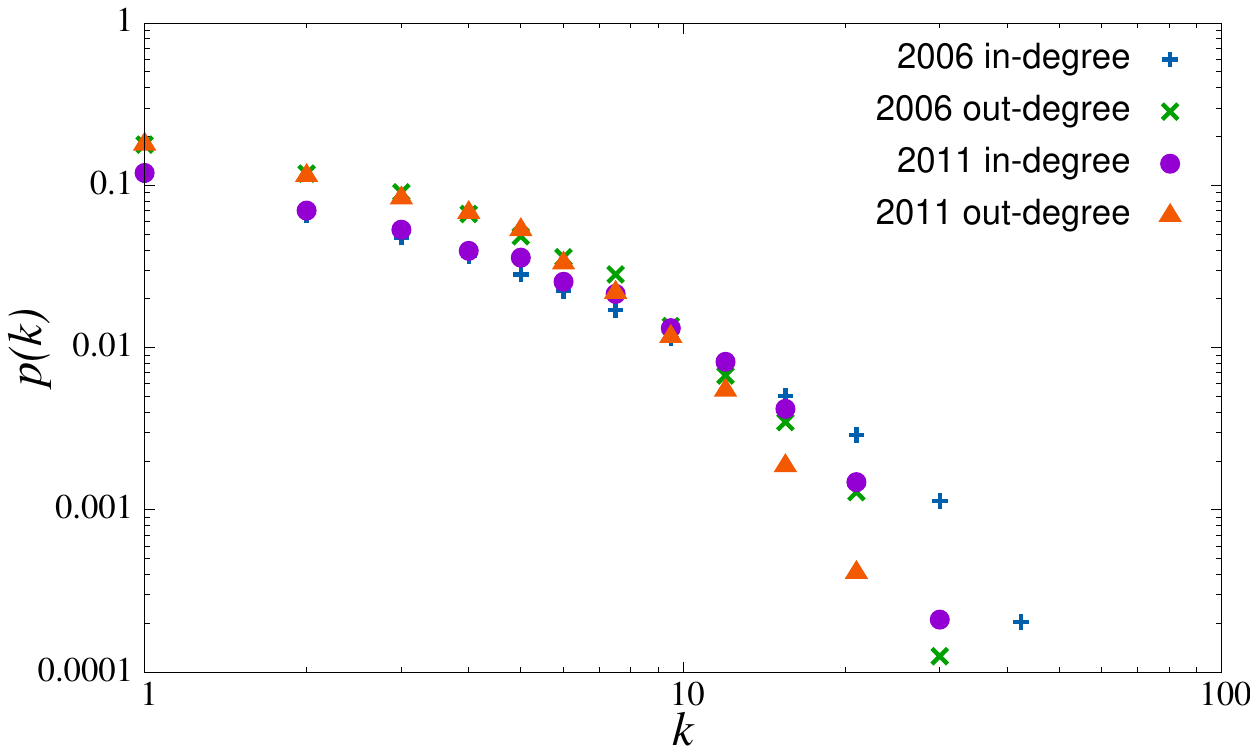}
\end{minipage}
\begin{minipage}[b]{0.49\linewidth}
\centering
\includegraphics[width=\linewidth, trim= 20 268 30 270, clip=true]{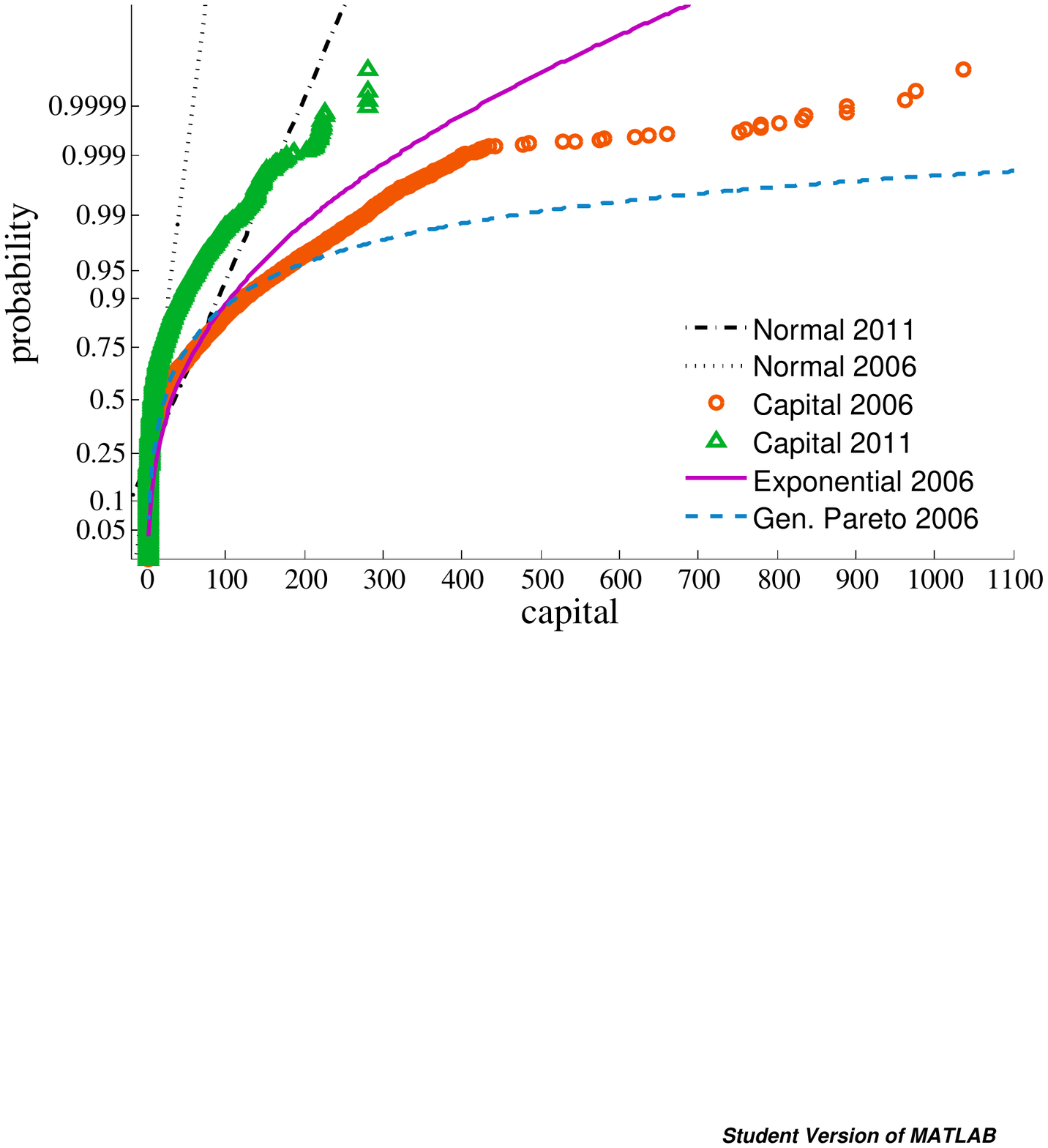}
\end{minipage}
\end{center}
\caption{Distribution of degree and capital}{\footnotesize Left: The distributions for in- and out-degree differ slightly. For both years the distribution is wider for the out-degree (lending) than for the in-degree (borrowing). The distributions have a tail and remind of a power-law with cutoff. Right: The (estimated) capital (in million euro) of banks is also very unevenly distributed. We show the CDF for the empirical distribution of capital for 2006 and 2011 together with a fitted Normal distribution. For 2006 we also show a fitted exponential and generalized Pareto distribution. (For this illustration we plot the capital for all banks and all trading days.) }
\label{fig:degree}
\end{figure*}

Put differently, this means that we assume some recovery for defaults in the successive rounds of bank failures. Defaulted banks will use their capital to repay interbank liabilities as good as they can, but they will not use any of the deposits. The exact design of this recovery rate might be debatable. We think that the assumption of some recovery is more realistic than the assumption of no recovery at all (which is implicitly assumed in some other studies), because it has a tendency to inflate domino effects.

\section{Simulation results}\label{sec:results}

\subsection{General results}

For the simulation we use daily empirical loan networks. We simulate cascades by iteratively choosing each bank as the possible source of a cascade, and we repeat this process for each day in the data set. For each single run we calculate if a cascade occurred, how many banks are affected, and how large the aggregated loss is. We can then related the magnitude of the cascade to the idiosyncratic properties of the banks, i.e., we can analyze in how far the size of the initial shock, the capital buffer, or the network structure influence the outcome.

The simulation with empirical data has some features that set it apart from simulations with random graphs. One feature is that the activity of the banks varies over time. In 2011 for example, the average number of banks that are active on one day is 70, the minimum is 38 and the maximum is 89 (2006 average: 128, min: 77, max: 144). The reason are mainly the bank holidays, which are not synchronized between countries. The average degree is more stable, for 2011 its fluctuations mostly stay in the band between 4 and 5. The degree distribution is somewhat close to a power law, the number of nodes is of course too small for a precise classification. Figure \ref{fig:degree} shows that the maximum of the out-degree (borrowing) is slightly larger than the in-degree (lending). This is caused by the fact that some larger banks serve as a kind of hub for the provision of liquidity, which leads to a slight asymmetry in the market.

Another feature of the data is that we can have more than one component. This means that parts of the network can be separated from the giant component. This can happen, when for example one small bank is borrowing from only one other bank, which itself has no other connections. This feature arises naturally from the strong heterogeneity in the size of banks. We checked that the large majority of banks are in one large component. In 2006 (2011) the average number of components was 1.7 (2.3).

\subsection{Influence of the size of the shock on the cascade}

\begin{figure*}[t]
\begin{center}
\begin{minipage}[b]{0.49\linewidth}
\centering
\includegraphics[width=\linewidth]{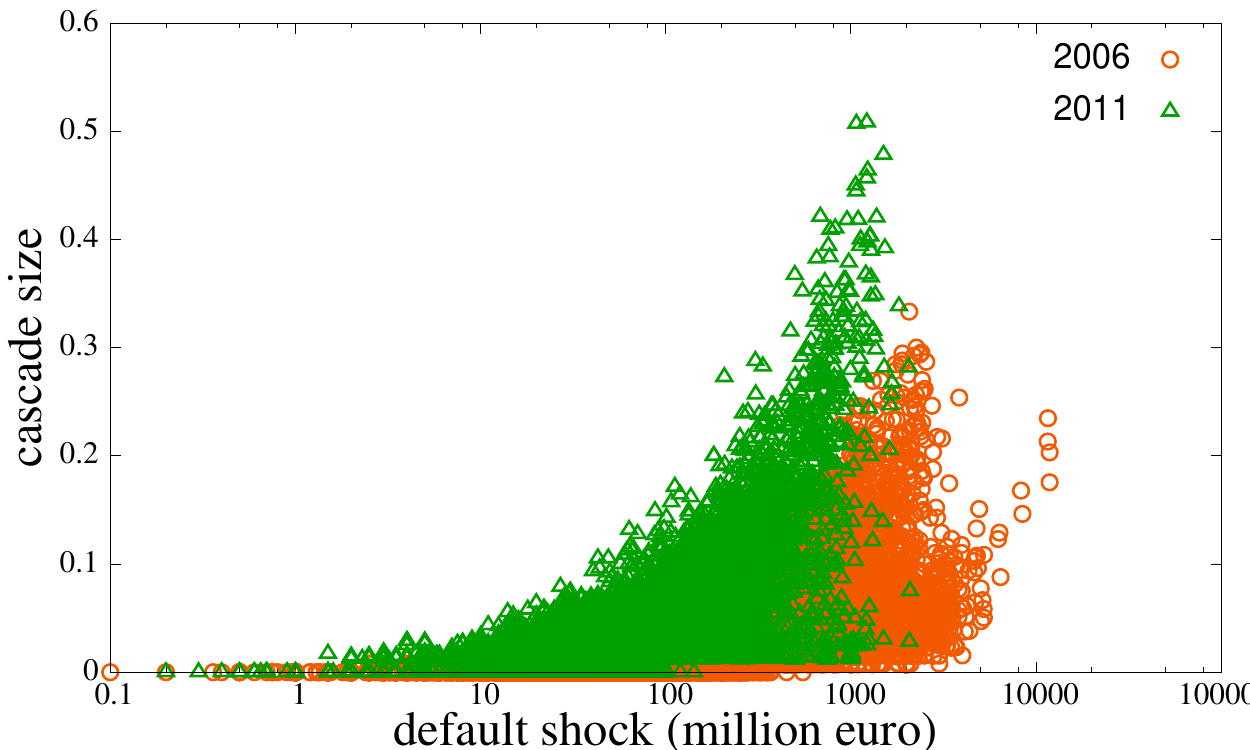}
\end{minipage}
\begin{minipage}[b]{0.49\linewidth}
\centering
\includegraphics[width=\linewidth]{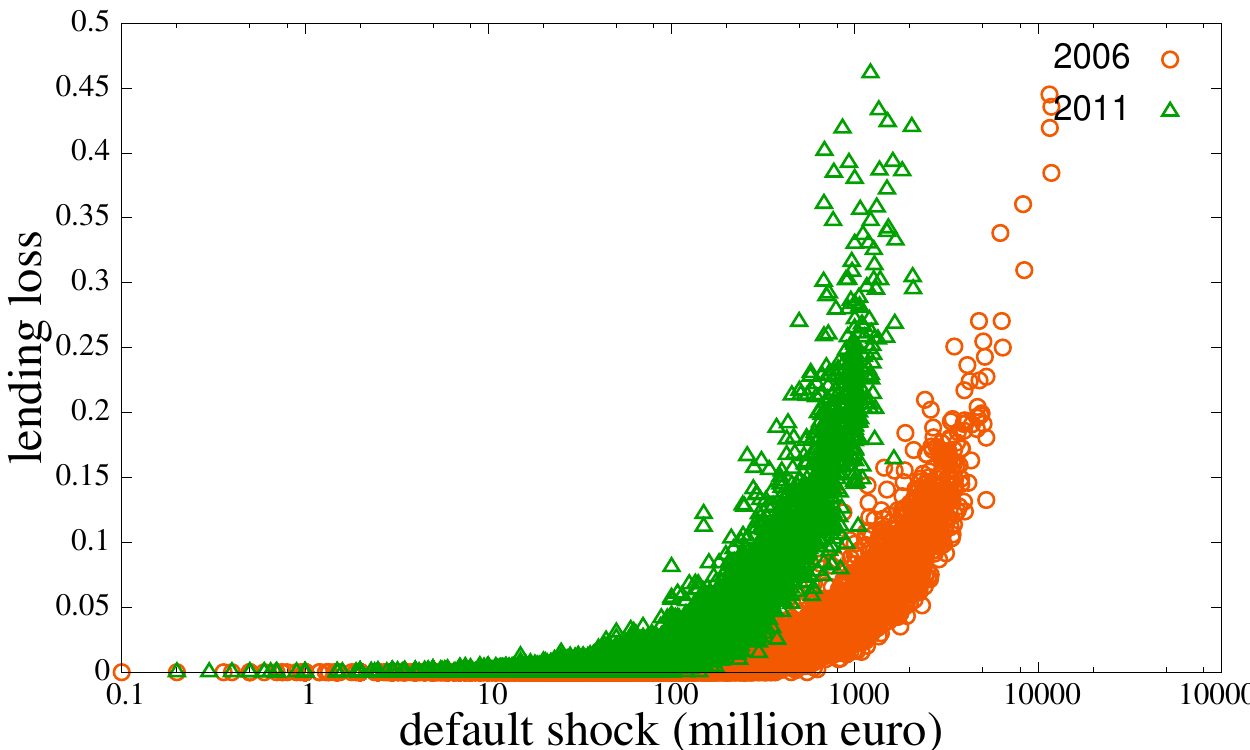}
\end{minipage}
\end{center}
\caption{Influence of shock on cascade size}{\footnotesize In these scatter plots of simulation results, the shock size is measured by the amount of failed loans of the bank that initially defaults. The left panel shows the resulting cascade measured by the share of banks that default as a consequence of this shock. In the right panel the cascade is measured by the share of interbank loans that fail as a consequence of the initial default. }
\label{fig:cascadesize}
\end{figure*}

Whether the initial default of a bank leads to significant losses to the entire system depends heavily on the size of the initial shock. For 2006 we analyze 25,914 hypothetical initial defaults, in 13,206 cases (50.1 \%) this default triggers the successive default of at least one other bank. In 2011 the number of transactions and also banks is much lower, thus we investigate only 13,614 cases of hypothetical initial default, 8,385 of which lead to at least one knock-on default of another bank (61.6 \%). The first impression from these numbers is that the interbank market has become more fragile. One reason is that banks share risks with fewer counterparts in 2011 \citep[see also][]{me}. However, it should be noted that in the simulations the ratio of interbank market activity to capital is kept constant. Even though the interbank network might be backed by more capital after the financial crisis, we can still conclude that the network is structurally more prone to cascade effects in 2011 than it was in 2006.

Figure \ref{fig:cascadesize} shows that the cascade size in our simulation depends on the size of the shock from the initial defaulter. In network science often a threshold is defined after which knock-on effects are regarded as a cascade, e.g., when more than 5\% of the nodes are affected. If such a measure is used in 2006 (2011) we have 904 (1782) cases, 3.5\% (13.1\%), in which a cascade happens. The pure number of affected nodes is not necessarily the best measure to quantify the cascade, because the banks which they represent can be of very different size. Hence, the right panel shows the cascade size measured by the fraction of total lending losses. The difference is that now cases in which high initial shocks lead to the default of only few but probably big counterparts are more accurately displayed, the same holds for small initial defaults that only trigger further small banks. When we apply the same 5\% threshold to this measure, we find that the difference between the years appears smaller, since in 2006 4.2\% of initial defaults result in a cascade compared to 8.5\% in 2011.\footnote{The difference in the cascade size between 2006 and 2011 is not caused by a sudden change. We simulated the cascade size for all remaining years, and found that the average cascade size gradually increases, while the average degree of the network steadily decreases throughout the years.}

\subsection{Sensitivity to parameters}\label{sec:sensitivity}

\begin{figure*}
\begin{center}
\includegraphics[width=0.8\linewidth, trim= 0 280 0 270, clip=true]{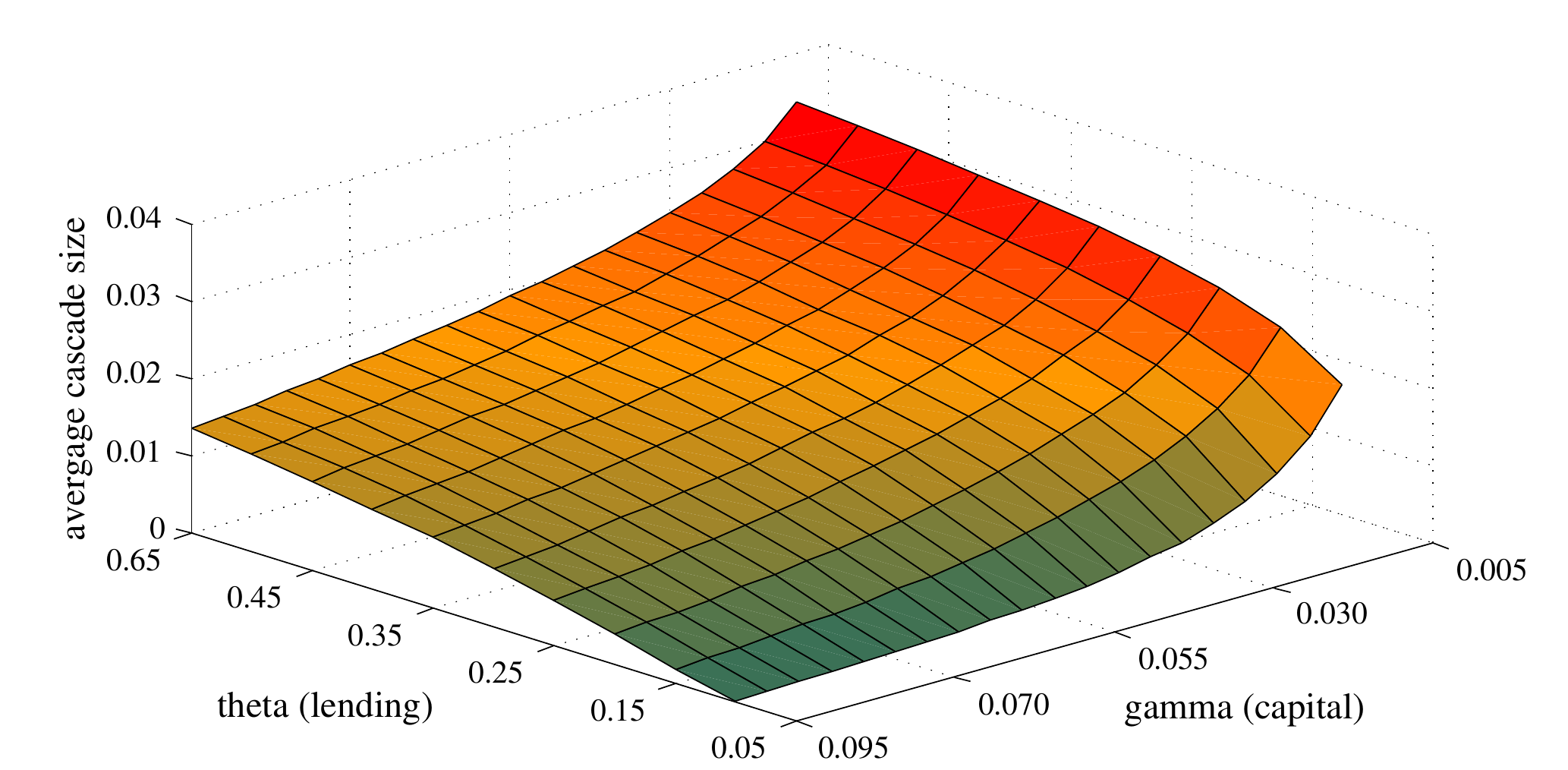}
\end{center}
\caption{Influence of gamma and theta}{\footnotesize A reduction in the ratio of capital to total assets $\gamma$ has a non-linear effect on the average cascade size, because at low levels of capital the chance of domino effects increases. Increasing the ratio of interbank assets to total assets $\theta$ leads to an almost linear increase of the average cascade size.}
\label{fig:parameters}
\end{figure*}

While the linkages in the interbank market for our simulation are derived from empirical data, and can thus be regarded as representative for the interbank assets, the remainder of the bank's characteristics are only approximated. The absolute numbers of certain balance sheet items are in fact not important for the results that we obtain. The parameters $\theta$ and $\gamma$, which describe the share of interbank assets and the ratio of capital to total assets, are however very important. Figure \ref{fig:parameters} shows their influence on the average cascade size. The important difference between them is that the cascade size increases only slowly once $\theta$ passes some upper bound, while lowering $\gamma$ leads to a critical point after that defaults sky-rock. Although these results are qualitatively similar to those of \cite{nier} for the Bernoulli random graph, heterogeneity of banks and network structure lead to more realistic results. Different from \cite{nier} we observe dependency from $\theta$ for a wider range of values, and a continuous increase in the cascade size for $\gamma$ (as a result of the network structure and the larger number of banks). As discussed in section 2.1, our choice for the parameter values for the baseline scenario are an approximation based on empirical findings. However, the qualitative results presented in the remainder of section 3 do not critically depend on the exact choice of these values.

\begin{figure*}
\begin{center}
\includegraphics[width= 0.8\linewidth, trim= 30 310 20 280, clip=true]{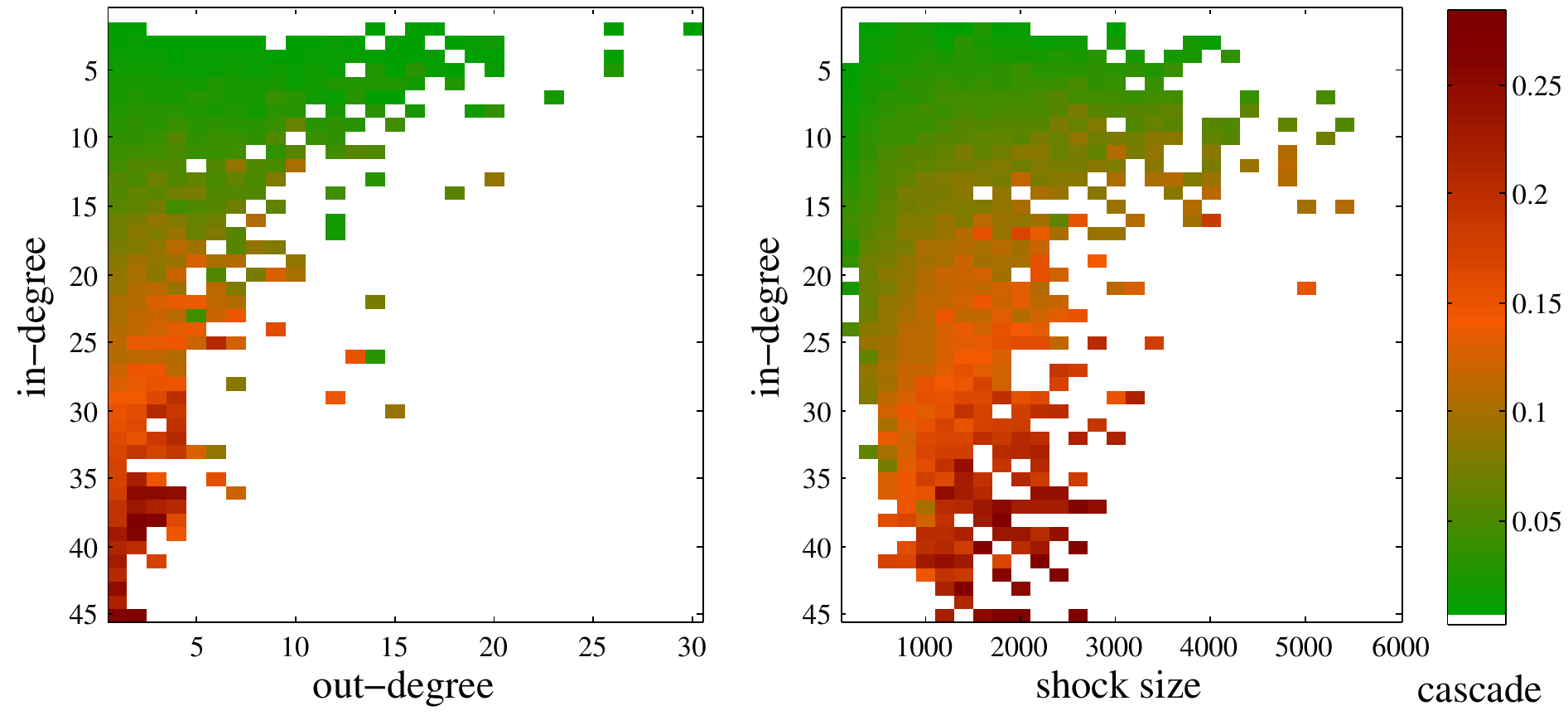}
\end{center}
\caption{Degree, shock size, and cascade size caused by the initially defaulted bank}{\footnotesize The heat maps show the average cascade when a bank with a specific degree defaults in 2006. The left panel shows that the cascade size increases with the in-degree (borrowing) of the defaulted bank. The right panel shows the relationship of in-degree and the shock size, i.e. the amount of loans that are not re-payed by the defaulted bank. We observe that large shocks do not necessarily lead to the largest cascades; medium sized shocks among well connected banks can affect even more banks.}
\label{fig:initial_degree}
\end{figure*}

\subsection{Influence of node connectivity on the cascade}

\begin{figure}[t]
\begin{center}
\includegraphics[width=\linewidth]{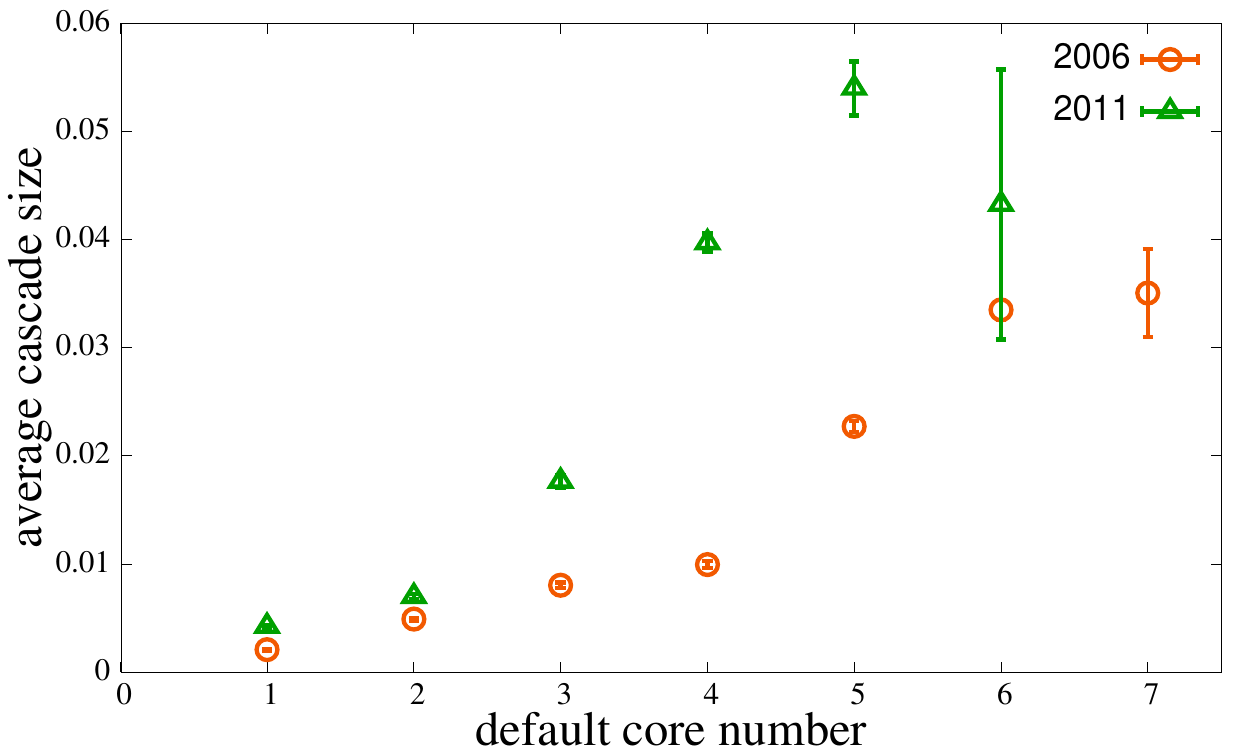}
\end{center}
\caption{Core number of initial defaulting bank}{\footnotesize The average cascade size increases with the core number of the initially defaulted bank. (The vertical bar shows the standard error.)}
\label{fig:initial_bc}
\end{figure}
In the following we analyze the influence of individual node characteristics on the cascade size and on the chance of survival of banks after a shock. We start with an analysis of the initially defaulting bank. 

We have already seen that the cascade size depends on the size of the initial shock. However, besides the pure size of the shock the connectivity of a bank and the network structure play a role \citep[see also][]{debtrank}. The left panel of Figure \ref{fig:initial_degree} shows the average cascade size dependent on the in- and out degree of the bank that we choose to default. Interestingly, we observe only few banks that have large in- and out-degree. The cascade size increases with the in-degree (borrowing of the defaulted bank), as can be expected. More interesting is to disentangle the influence of in-degree and shock size. The right panel shows that even medium sized shocks can cause large cascades, if banks are well connected. It has to be noted that some of the larger shocks in our data set stem from large international banks, which connections to other global players outside the Italian E-mid platform cannot be accounted for in this simulation. However, the analysis clearly shows that for any given shock size greater connectivity leads to larger cascades, i.e., a larger number of defaulted banks.

\begin{figure*}
\begin{center}
\includegraphics[width= \linewidth, trim= 27 330 25 320, clip=true]{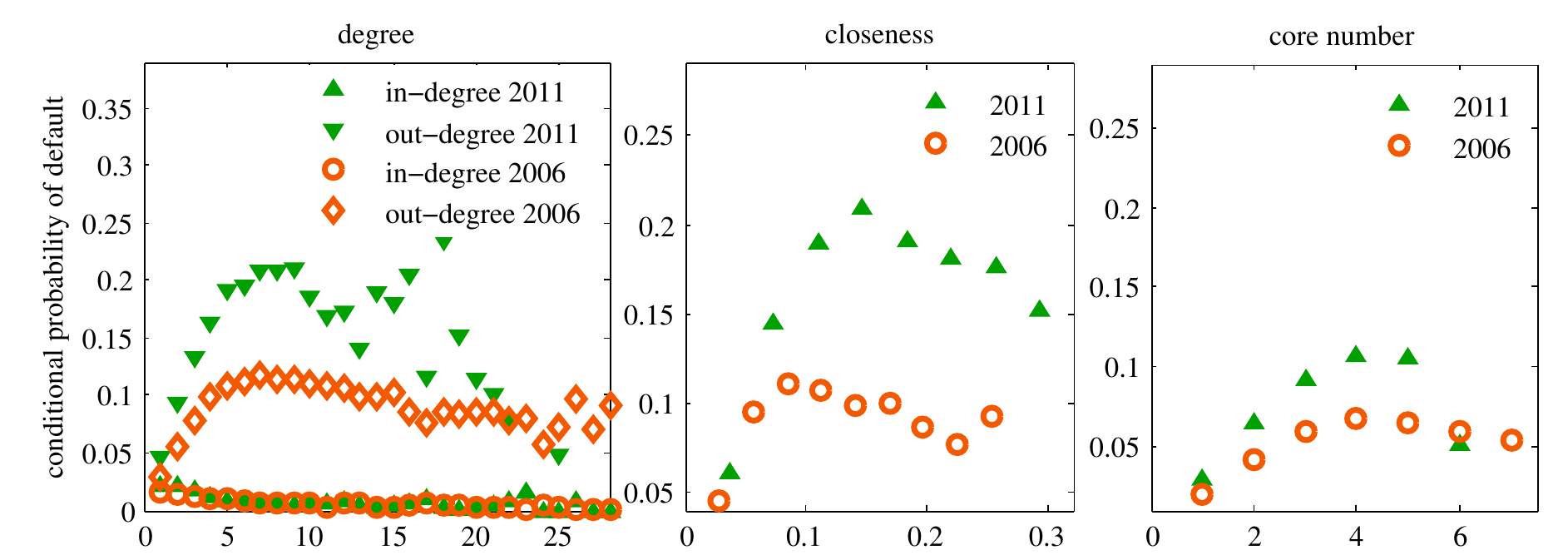}
\end{center}
\caption{Degree, closeness,  core number, conditional default probability}{\footnotesize Left panel: The conditional probability of a bank to default as part of a chain reaction decreases with the in-degree (borrowing). For the out-degree (lending) we observe a hump-shaped relationship. Middle panel: Banks with an intermediate closeness are the ones most prone to default in our simulation. Right panel: Also for the core number we observe a hump-shaped relationship with the probability to default.}
\label{fig:vulnerable}
\end{figure*}

A related measure for centrality is the core number of a node. A k-core is defined as the largest subgraph which nodes have at least $k$ connections \citep[see][]{kcore}. Hence, if a node has a high core number, this shows that it has not only a high degree, but that it is also connected to nodes with a similar high degree. Figure \ref{fig:initial_bc} shows that in fact the cascade size increases with the core number. Previous studies on disease spreading models have also shown this effect of coreness on the disease outbreak size \citep[see, e.g.,][]{spreaders}.

After analyzing which bank is the most ``dangerous'' when it defaults, we can now turn to the analysis of the vulnerability of banks to this default. Hence, we look at measures of node centrality to determine which characteristics (besides the shock size) determine the probability of a bank to default at any stage of a cascade process. We start with the degree, which is reported in the left panel of Figure \ref{fig:vulnerable}. For both years, 2006 and 2011, the conditional probability to default is decreasing with the in-degree (borrowing), because banks that borrow a lot are less vulnerable (and mostly lend less than they borrow). The results for the out-degree are more interesting. In general, we would expect that the default probability decreases with degree, since shocks are wider distributed among banks. However, for 2006 we observe a hump-shaped curve, the probability to default increases for a degree of up to 7, before it slowly decreases for banks with more links. For 2011, where the network is smaller, we barely reach the point where the probability decreases (the results for degree > 15 are noisy due to the small sample). We can infer that this sample contains a lot of banks that in terms of risk sharing are too small. Some of these might in fact be part of an affiliate group of banks.

The analysis of some centrality measures helps us to understand these patterns. The node closeness describes how easily a node can be reached from any other one in the network \citep[see, e.g., ][]{freeman}. The middle panel of Figure \ref{fig:vulnerable} shows that nodes (banks) that have low closeness, and are thus difficult to reach for a shock, have a low probability to default. The probability increases for values of up to 0.26 and then decreases for nodes with a higher closeness. Put differently, for the connectivity of banks there is a threshold after which the likelihood to be hit by a shock is out-weight by the benefits from spreading exposures.
We obtain similar results from an analysis of the default probability dependent on the core number. The probability to default is highest for banks with a core number of 4-5, and decreases for values higher than this.

Our analysis emphasizes that degree, centrality, and coreness are crucially related to the banks' ability to withstand shocks. Models with random interconnections between banks cannot reproduce all these findings. Therefore they demand further investigation. This is evenmore true for the unclear influence of the heterogeneity in the size of banks. In the next section, we compare the results from the empirical cascade simulation with the results from null models to disentangle the effect of network structure and heterogeneity in the size of banks.

\section{Null models}\label{sec:null}

\begin{figure*}
\begin{center}
\includegraphics[width= 0.85\linewidth, trim= 0 40 20 55, clip=true]{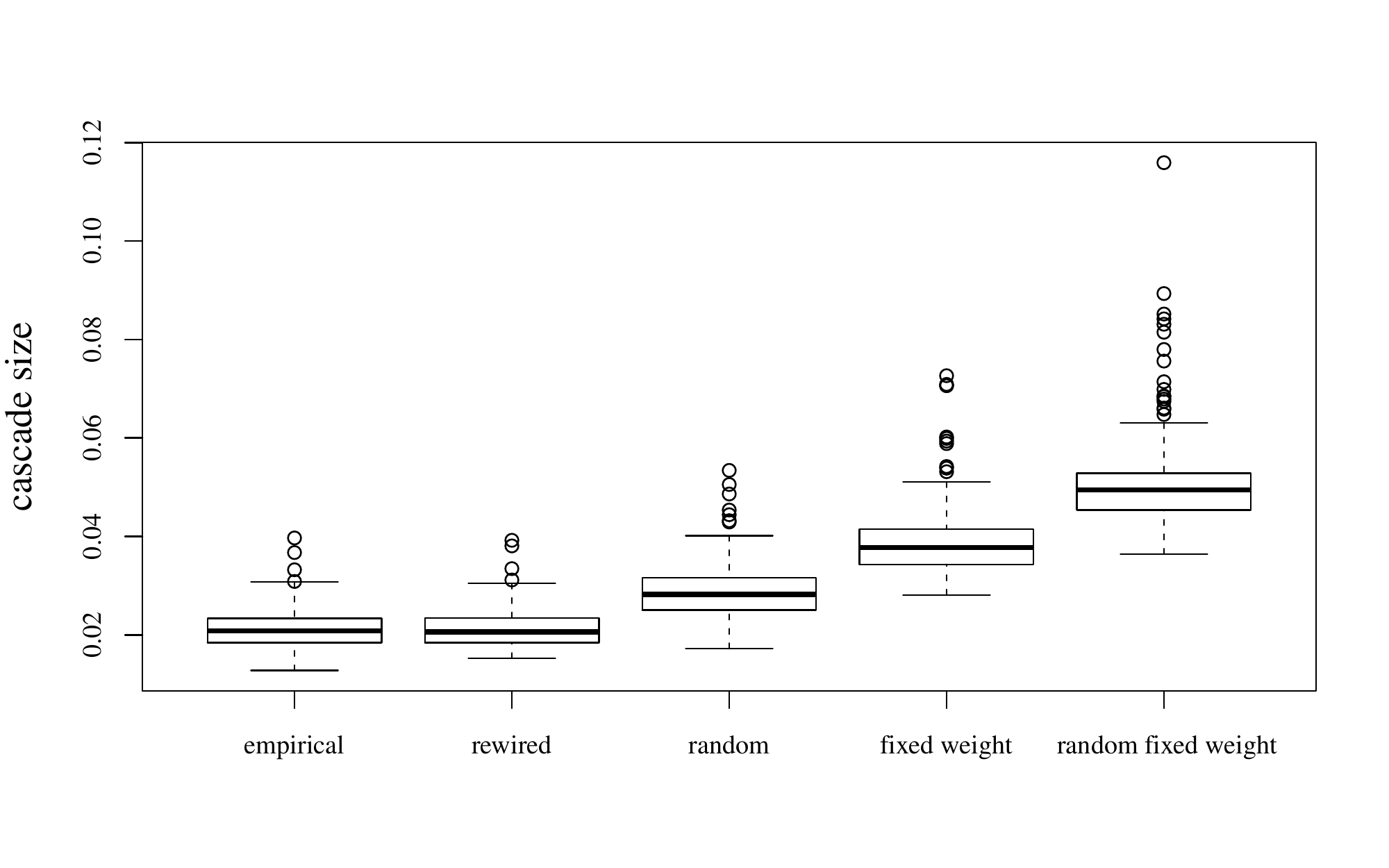}
\end{center}
\caption{Cascade size for different null models}{\footnotesize The box-plots show the average daily cascade size for the empirical model (left) compared to the null models described in this section for 2011. The rewired and random null model lead to a slightly higher cascade size. The null models that randomize both, size distribution and links, lead to much larger cascades than the empirical model.}
\label{fig:null}
\end{figure*}

The interbank network has several interesting structural features that are likely to have a significant impact on its stability. Among those are the heterogeneity in the bank size, the directionality and assortativity of links, and the heterogeneity in the link weights. In order to investigate the role of each of these features we need to derive a benchmark; this means that we have to derive the average cascade size in a network where these features are excluded or randomized. Such a benchmark network is called a null model. Null models are an established tool to separate ``natural'' network effects from influences that are caused by the actual degree sequence of a network, i.e., behavior that is likely to be strategic and planned, or the outcome of the environment. Null models have been studied in ecological networks \citep[see, e.g.,][]{vaz}, protein networks \citep[see, e.g.,][]{maslov}, commmunication networks \citep[see, e.g.,][]{karsai}, face-to-face interaction networks \citep[see, e.g.,][]{isella} and recently in trade networks \citep[see][]{fagiolo}.

Hence, we compare the stability of the model based on the empirical network of the interbank market with the stability of different null models. In all cases we preserve the number of nodes, the number of links, and the directionality of the network. Specifically, we generate the following null models:

In the \emph{rewired} null model we preserve the degree of each nodes, and the weight of each link (loan size). The algorithm randomly selects pairs of links with similar size and rewires (switches) the counterparts of these links, if the rewired links do not yet exist. If the rewired links are already present in the network, the algorithm picks another pair with similar weights. The banks' balance sheets are adjusted to be consistent with the slight changes in the bank's interbank market activity. This null model is similar to one proposed by \cite{maslov}. The model should show in how far the non-randomness of connectivity influences the stability of the network, given the very heterogeneous link weights.

In the \emph{random} (Erd\"os-R\'{e}nyi) null model we start out with the empirical set of loan contracts and generate random links between the banks using the empirical loan contract weights. This null model destroys the topological structure of the network, but preserves the daily interbank volume. The size of the banks, their capital, and their degree are also randomized. The banks' balance sheets are again adjusted to be consistent with interbank market activity. 

The \emph{fixed weight} null model preserves the number of links for all banks, but assigns constant weights to all of them. For each day, the constant weight is calculated by dividing the total lending by the number of contracts on that day. Hence, this null model preserves the degree distribution, but makes the link weights homogenous. Accordingly, the size of banks (and their capital) is proportional to the degree.  

Finally, we assume a \emph{random network with fixed weights}. In this null model the link weights are constant, as described in the previous model. Moreover, we randomize the connectivity of the nodes like in the random model. This model has the least similarity with the empirical network; it also has the largest deviation in the resulting average cascade size.

Figure \ref{fig:null} illustrates the size of the cascade in different null models compared to the results from the empirical network model of 2011 with its baseline parametrization.\footnote{The results are very similar for the empirical network from 2006.} For the rewired model the average cascade size is comparable to that from the empirical network. The random null model, which destroys the degree connectivity but keeps the weights, results in a 1.4 times higher cascade size. The fixed weight model leads to an even 1.9 times higher average cascade. The random fixed weighted model results in 2.5 times more defaults than the empirical model.
The analysis of the null models reveals that simulating bank networks assuming homogeneity in the bank and loan size dramatically overestimates its fragility. In comparison, the ``error'' from ignoring the size distributions of loans and banks seems to be much greater than assuming a wrong distribution of links and degree.

In section \ref{sec:sensitivity} we have shown that the cascade size depends on the lending activity and on the average amount of capital, i.e. the choice of the parameters $\gamma$ and $\theta$. Similarly we can analyze the sensitivity of the relative differences in the cascade size for the different null models. To reduce the dimensionality of the analysis we will only look at a variation of the ratio of capital to total assets $\gamma$. This is without loss of generality, since the system could also be described by one single parameter, the ratio of lending to capital (the system has one free parameter). 

\begin{figure*}
\begin{center}
\includegraphics[width= 0.85\linewidth, trim= 0 270 0 290, clip=true]{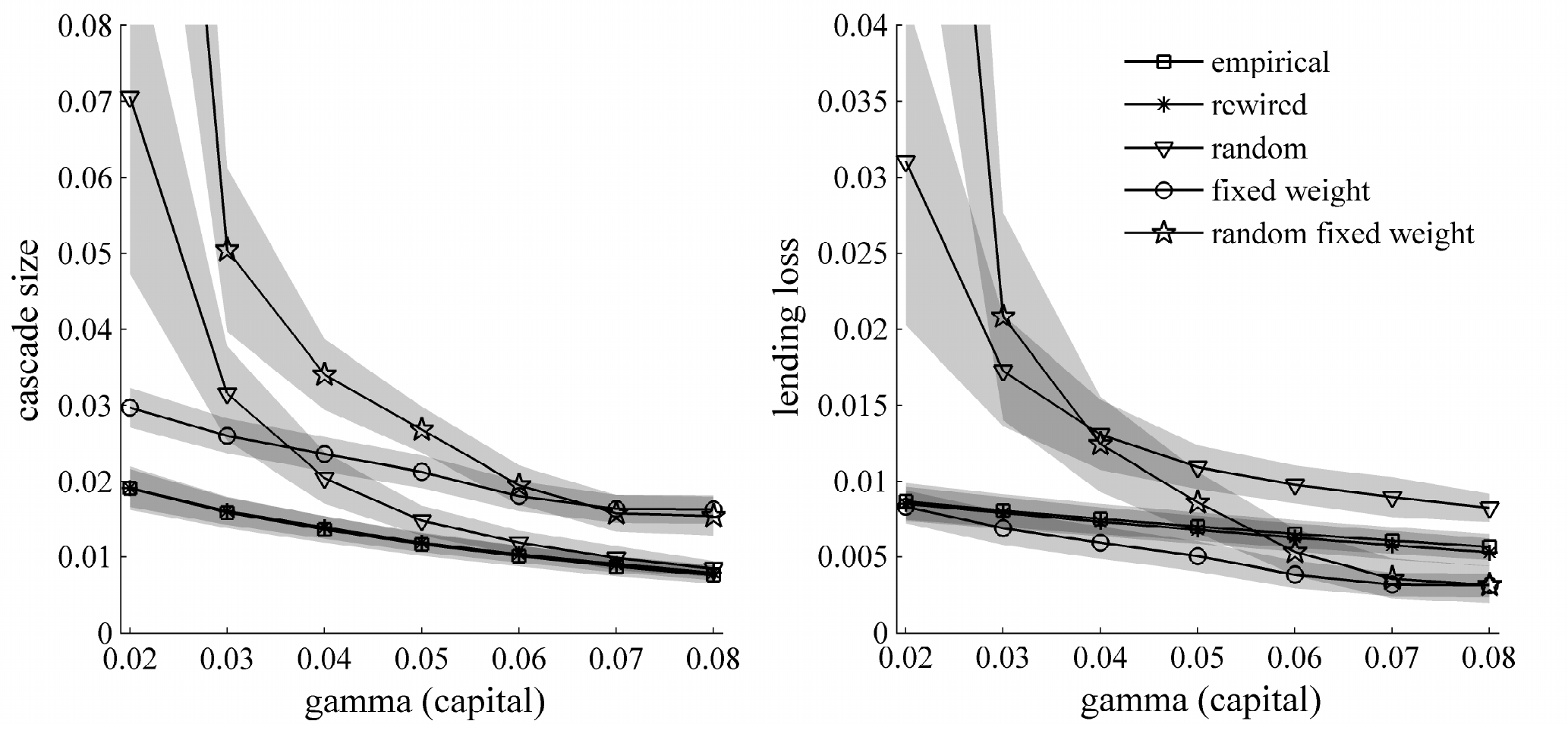}
\end{center}
\caption{Deviations for the null models for different levels of gamma}{\footnotesize Left: Null models with randomized network structure show a exaggerated increase in cascade size for systems with low capital compared with the empirical model. Right: When the cascade size is measured by the lending loss the default of smaller banks appears less severe in the fixed weight and random fixed weight model, the defaults in the random model show relatively stronger than in the empirical model. The shaded areas show standard errors.}
\label{fig:null_all}
\end{figure*}

The left panel of
figure \ref{fig:null_all} shows the average cascade size for all five models depended on $\gamma$ (or, more generally speaking, for different levels of fragility). The deviation of the null models from the empirical model differ most when $\gamma$ is decreased to values much lower than in the baseline parametrization. While the fixed weight null model seems to result in a constant overestimation of the cascade size, the two models with a randomized network structure show exponentially increasing deviations in the cascade size for low values of $\gamma$. For the fixed weight model we observe a greater likelihood of chains of defaults of small banks than in the empirical model. In this case the fixed weights lead to the creation of many similar sized small banks, which all cannot withstand defaults of slightly larger banks. The increasing deviation of the models with the randomized network structure stems from the micro-structure of the overnight loan market. In the short run most of the banks are predominantly active on one side of the market, i.e., excess liquidity or demand leads to the fact that they focus on either borrowing or lending. Since a bank can only pass shock to other banks if it is active on both sides of the market, the effect of randomization of the network structure is especially large when the system is rather unstable (low $\gamma$) and chains of defaults gain importance versus direct effects of an initial default. The rewired model shows cascade sizes that are almost indistinguishable from the empirical model, which leads to the conclusion that the in- and out-degree distributions together with bank size is a sufficient description of the system.

Alternatively we can again look at the lending loss as a measure of the severity of the cascade. The results for the different models are presented in the right panel of figure \ref{fig:null_all}. Although the absolute numbers are smaller here, the qualitative results are similar. In comparison with the left panel, the models which use fixed link weights generate less lending loss for relatively high values of $\gamma$ than the empirical model. This is caused by the default of banks that appear smaller than in the empirical model. This effect of the fixed link weights vanishes once $\gamma$ becomes relatively small. In these cases the homogeneity in the size distribution leads to more chains of defaults.

\section{Conclusions}\label{sec:conclusions}

In this paper we analyzed the likelihood of cascades in an empirical interbank loan market. 
The novel aspect of the paper is that we assume heterogeneity in the size of banks and loans, which are derived directly and indirectly from empirical data.
Our results indicate that the cascade size depends not only on the size of the initial shock, but also on the in-degree of the defaulter. Moreover, the position of the initial defaulter within the network, e.g. its centrality and coreness, plays an important role for its ability to start a cascade.

We found that the fragility of this market, given our model, has increased significantly from 2006 until 2011. Our analysis of the vulnerability of banks shows that for 2006, in line with theoretical results on risk sharing, the probability to default decreases with the out-degree. For 2011 however, we do not observe the same pattern, instead there is strong positive correlation between the centrality and vulnerability of banks. It seems likely that the increase in vulnerability is linked to a change in lending behavior, caused by the lost trust in many borrowers, and that this change feeds back as a negative influence on overall system stability.

To analyze specific aspects of the network structure on the cascade process, we made use of different null models. The results show that the empirical network has features, which cannot be captured by randomized network models. Furthermore, we show that assuming homogeneity for bank size and link weights can lead to an overestimation of the fragility of the system.

We believe that this study is an attempt to model more realistic financial networks. Further research should focus on the influence of temporal aspects of the interbank network on cascades of defaults. Another research direction is to model multiplex networks that incorporate different financial products and markets.

\section*{Acknowledgments}

The authors thank Petter Holme, Martin Rosvall, and Thomas Lux for helpful discussions. FK thanks the Swedish Research Council. MR thanks the Leibniz Association for partial funding of this project.

\end{document}